\documentclass[draft]{agujournal2019}
\usepackage{url} 
\usepackage{lineno}
\usepackage[inline]{trackchanges} 
\usepackage{soul}
\nolinenumbers

\newcommand{\degr}{$^{\circ}$ }
\draftfalse

\journalname{Journal of Advances in Modeling Earth Systems (JAMES)}

\begin{document}

%
%

\title{Data-driven medium-range weather prediction with a Resnet pretrained on climate simulations: A new model for WeatherBench}

%
%




\authors{Stephan Rasp\affil{1}\thanks{Now at: ClimateAi}, and Nils Thuerey\affil{1}}


\affiliation{1}{Department of Informatics, Technical University of Munich, Germany}




\correspondingauthor{Stephan Rasp}{stephan@climate.ai}




\begin{keypoints}
\item A large convolutional neural network is trained for the WeatherBench challenge.
\item Pretraining on climate model data improves skill and prevents overfitting.
\item The model sets a new state-of-the art for data-driven medium-range forecasting.
\end{keypoints}

%
%

%
%


\begin{abstract}
Numerical weather prediction has traditionally been based on physical models of the atmosphere. Recently, however, the rise of deep learning has created increased interest in purely data-driven medium-range weather forecasting with first studies exploring the feasibility of such an approach. To accelerate progress in this area, the WeatherBench benchmark challenge was defined. Here, we train a deep residual convolutional neural network (Resnet) to predict geopotential, temperature and precipitation at 5.625 degree resolution up to 5 days ahead. To avoid overfitting and improve forecast skill, we pretrain the model using historical climate model output before fine-tuning on reanalysis data. The resulting forecasts outperform previous submissions to WeatherBench and are comparable in skill to a physical baseline at similar resolution. We also analyze how the neural network creates its predictions and find that, with some exceptions, it is compatible with physical reasoning. Finally, we perform scaling experiments to estimate the potential skill of data-driven approaches at higher resolutions.
\end{abstract}

\section*{Plain Language Summary}
Weather forecasts are created by running hugely complex computer simulations that encapsulate our knowledge of how the atmosphere works. This approach has served us well but is there a different way? The paradigm of machine learning proposes learning an algorithm from data rather than building it from physical principles. For several areas like computer vision and natural language processing this has worked exceedingly well, so it just makes sense to try it as well for weather forecasting. This paper presents the latest attempt at training a machine learning weather forecasting model. It is shown that the learned model produces reasonable forecasts, approximately on par with traditional models run on much lower resolution. However, there is still a large gap to current state-of-the-art high-resolution weather models that is unlikely to be closed with a purely data-driven approach because not enough training data exists.

%
%

%


%
%
%
%

\section{Introduction}
Current numerical weather prediction (NWP) is based on physical models of the atmosphere, and the ocean for longer forecast times, in which the governing equations are discretized and sub-grid processes are parameterized \cite{Kalnay2003}. Continued refinement of these models along with increasing computing power and better observations to create initial conditions has led to steady increases in forecast skill over the last four decades \cite{Bauer2015a}. The improvements in the model components and the tuning of free parameters is, in a large majority of cases, guided by scientific expertise rather than using a statistical method \cite{Hourdin2017}. In the current operational weather forecasting chain, the only component that includes a learning algorithm is post-processing, the correction of statistical errors from NWP output. Most commonly, post-processing is done using simple linear techniques (model output statistics; MOS) but in recent years more modern machine learning (ML) techniques, such as random forests and neural networks, have been explored \cite{Taillardat2016, McGovern2017, Rasp2018d, Gronquist2020}. 

With the apparent successes of deep learning in modeling high-dimensional data in other domains such as computer vision and natural language processing, a natural question to ask is whether numerical weather models can also be learned purely from data. This question sparked some debate after initial studies \cite{Dueben2018,Scher2018,Scher2019,Weyn2019} showed the general feasibility of such an approach for medium-range weather forecasting. In particular, some researchers were sceptical whether the complex physics described by systems of partial differential equations could be encoded in a neural network. 

To answer this question \citeA{Rasp2020a} defined a benchmark challenge for data-driven medium-range weather forecasting called WeatherBench. Specifically the challenge is to predict 500\,hPa geopotential (Z500), 850\,hPa temperature (T850), 2-meter temperature (T2M) and 6-hourly accumulated precipitation (PR) up to 5 days ahead. Here, we train a large neural network for this task. Section~\ref{sec:methods} describes the dataset and the neural network setup. In section~\ref{sec:results} the results for the WeatherBench benchmark are presented and discussed. Section~\ref{sec:sensitivity} includes several sensitivity experiments followed by an attempt to interpret the neural network's predictions in section~\ref{sec:interpretability}. Finally, we discus the results in section~\ref{sec:conclusion}.

\section{Materials and Methods}
\label{sec:methods}
\subsection{Data, Evaluation and Baselines}
The data is provided by the WeatherBench challenge. A full description can be found in \citeA{Rasp2020a} and the latest version of the data is available at \url{https://github.com/pangeo-data/WeatherBench}. WeatherBench contains regridded ERA5 \cite{Hersbach2020} data from 1979 to 2018 at hourly resolution. In addition, we use climate model simulations to pretrain our simulations as described below. For this we downloaded a historical simulation from the CMIP6 archive \cite{Eyring2016}. Specifically, we picked the MPI-ESM-HR model since it was one of the only models for which the data was saved at vertical resolution to match the ERA5 data. The temporal resolution is six hours. The regridded climate model data are also available on the WeatherBench data repository. In addition to the data, WeatherBench defines the evaluation metrics. The area-weighted RMSE and ACC are used for evaluating 500\,hPa geopotential (Z500), 850\,hPa temperature (T850), 2-meter temperature (T2M) and 6-hourly accumulated precipitation (PR) at 3 and 5 days lead time. The area-weighted RMSE is defined as
\begin{equation}
\label{eq:rmse}
    \mathrm{RMSE} = \frac{1}{N_{\mathrm{forecasts}}} \sum_i^{{N_{\mathrm{forecasts}}}} \sqrt{\frac{1}{N_{\mathrm{lat}} N_{\mathrm{lon}}} \sum_j^{N_{\mathrm{lat}}} \sum_k^{N_{\mathrm{lon}}} L(j) (f_{i, j, k} - t_{i, j, k})^2} 
\end{equation}
where $f$ is the model forecast and $t$ is the ERA5 truth. $L(j)$ is the latitude weighting factor for the latitude at the $j$th latitude index:
\begin{equation}
    L(j) = \frac{\cos( \mathrm{lat}(j))}{\frac{1}{N_{\mathrm{lat}}} \sum_j^{N_{\mathrm{lat}}} \cos( \mathrm{lat}(j)) }
\end{equation}. The definition of the ACC can be found in the appendix.

Further, WeatherBench contains several baselines from physical models: the operational Integrated Forecasting System (IFS) of the European Center for Medium-range Weather Forecasting (ECMWF), the current state-of-the-art in NWP, which currently runs at 10\,km horizontal resolution with 137 vertical levels; and the same model run at two lower resolutions, T42 (approximately 2.8\degr or 310\,km at the equator) with 62 vertical levels and T63 (approximately 1.9\degr or 210\,km at the equator) with 137 vertical levels. For an exact definition of the evaluation metrics and the initialization of the baseline models, refer to \citeA{Rasp2020a}. As an additional baseline, here we include the work by \citeA{Weyn2020} who trained an neural network to predict Z500 and T850. Their model is iterative, i.e. it consists of a sequence of 6\,h forecasts. During training they also trained their neural network over two time steps (12\,h) to ensure stability for longer integrations. Further they mapped the latitude-longitude data to a cube-sphere grid with roughly 1.9\degr resolution to minimize the distortion during the convolution operations. Their model was trained on 40 years of ERA data.

\subsection{Data-driven forecasts using a pretrained Resnet}
There are three fundamental techniques for creating data-driven forecasts: direct, continuous and iterative. For direct forecasts, a separate model is trained directly for each desired forecast time. In continuous models, time is an additional input and a single model is trained to predict all forecast lead times (as in MetNet; \citeA{Sonderby2020}). Finally, iterative forecasts are created by training a direct model for a short forecast time (e.g. 6\,h) and then running the model several times using its own output from the previous iteration. As mentioned above, this is the approach taken by \citeA{Weyn2020}.

Here, we train direct and continuous models. Advantages and disadvantages of each technique will be discussed later. All models in this study use the same architecture (except in the network size scaling experiments). The basic structure is a fully convolutional Resnet \cite{He2015} with 19 residual blocks. Each residual block consists of two convolutional blocks, defined as [2D convolution $\to$ LeakyReLU $\to$ Batch normalization $\to$ Dropout],  after which the inputs to the residual layer are added to the current signal. The 2D convolutions inside the residual blocks have 128 channels with a kernel size of 3. All convolutions are periodic in longitude with zero padding in the latitude direction. For the first layer a simple convolutional block with 128 channels is used with a kernel size of 7 to increase the field of view. LeakyReLU is used with $\alpha = 0.3$. Weight decay of $1 \times 10^{-5}$ is used for all layers. Dropout is set to 0.1.

The inputs are geopotential, temperature, zonal and meridional wind and specific humidity at seven vertical levels (50, 250, 500, 600, 700, 850 and 925\,hPa), 2-meter temperature, 6-hourly accumulated precipitation, the top-of-atmosphere incoming solar radiation, all at the current time step $t$, $t-6$h and $t-12$h, and, finally three constant fields: the land-sea mask, orography and the latitude at each grid point. All fields were normalized by subtracting the mean and dividing by the standard deviation, with the exception of precipitation for which the mean was not subtracted to keep the lower bound at zero. Additionally, we log-transform of the precipitation to make the distribution less skewed ($\tilde{PR} = \ln(\epsilon + PR) - \ln(\epsilon)$) with $\epsilon = 0.001$. Subtracting the log of $\epsilon$ ensures that zero values remain zero. This transformation turns out to be crucial to prevent the network from simply predicting zeros. All variables, levels and time-steps were stacked to create an input signal with 114 channels. For the continuous forecast, in addition, we add a 32$\times$64 fields which contains the forecast time in hours divided by 100. During training, a random forecast time from 6 to 120 hours is drawn for each sample. Two separate sets of networks were trained, one to predict Z500, T850 and T2M and another one to predict TP. The reason for treating TP separately is that its distribution is significantly more skewed even after the log-transform compared to the other three variables. Predicting all four variables with a single network led to bad predictions for all variables. A loss scaling factor for TP be one potential solution but here we chose to simply treat it separately.

For our best models we first train our model using the 150 years of CMIP data described above. We then take the pretrained model and fine-tune it using the ERA data. We will also show results for models trained only with CMIP or ERA data. The loss function is the latitude-weighted mean squared error. The latitudes are weighted proportionally to the area of the grid boxes $\propto \cos \phi$. The Adam optimizer \cite{Kingma2014} is used with a batch size of 32 and an initial learning rate of $5 \times 10^{-5}$ for the ERA and CMIP only experiments. The learning rate was decreased twice by a factor of 5 when the validation loss has not decreased for two epochs. Early stopping on the validation loss was used to terminate training with a patience of 5 epochs. The training period for ERA was from 1979 to 2015, validation was done with a single year (2016). For fine-tuning the CMIP networks on ERA data, a lower initial learning rate of $5 \times 10^{-7}$ was chosen. Also note that for the pretrained model, no dropout was used as this led to better validation scores. For the direct approach we trained models for 6h, 1d, 3d and 5d forecast time. We used Tensorflow 2. Training a single model takes around one day on a GTX 2080 GPU.

\section{WeatherBench results}
\label{sec:results}

\begin{table}[th!]
\caption{RMSE for 3 and 5 days forecast time. All forecasts evaluated at 5.625\degr resolution. Best physical and data-driven methods are highlighted.}
\begin{tabular}{l||l|l|l|l}
 & \multicolumn{4}{c}{\hspace{0pt} Latitude-weighted RMSE (3 days / 5 days)} \\
\textbf{Model} & \textbf{Z500 [m$^2$ s$^{-2}$]} & \textbf{T850 [K]} & \textbf{T2M [K]} & \textbf{PR [mm]} \\ \hline \hline
Persistence & 936 / 1033 & 4.23 / 4.56 & 3.00 / 3.27 & 3.23 / 3.24 \\ \hline
Climatology & 1075 & 5.51  & 6.07 & 2.36 \\ \hline
Weekly climatology & 816 & 3.50  & 3.19 & \textbf{2.32} \\ \hline
IFS T42 & 489 / 743 & 3.09 / 3.83  & 3.21 / 3.69 & \\ \hline
IFS T63 & 268 / 463 & 1.85 / 2.52  & 2.04 / 2.44 & \\ \hline
Operational IFS & \textbf{154} / \textbf{334} & \textbf{1.36} / \textbf{2.03} & \textbf{1.35} / \textbf{1.77} & 2.36 / 2.59 \\ \hline \hline
Weyn et al. (2020) & 373 / 611 & 1.98 / 2.87 & & \\ \hline
Direct (ERA only) & 314 / 561 & 1.79 / 2.82 & 1.53 / 2.32 & \textbf{2.03} / 2.35 \\ \hline
Direct (CMIP only) & 323 / 561 & 2.09 / 2.82 & 1.90 / 2.32 & 2.30 / 2.39 \\ \hline
Direct (pretrained) & \textbf{268} / 523 & \textbf{1.65} / 2.52 & \textbf{1.42} / 2.03 & 2.16 / \textbf{2.30} \\ \hline
Continuous (ERA only) & 331 / 545 & 1.87 / 2.57 & 1.60 / 2.06 & 2.22 / 2.32 \\ \hline
Continuous (CMIP only) & 330 / 548 & 2.12 / 2.75 & 2.24 / 2.59 & 2.29 / 2.38 \\ \hline
Continuous (pretrained) & 284 / \textbf{499} & 1.72 / \textbf{2.41} & 1.48 / \textbf{1.92} & 2.23 / 2.33  \\ \hline

\end{tabular}
\label{tab:rmse}
\end{table}

\begin{figure}[th!]
 \includegraphics[width=\linewidth]{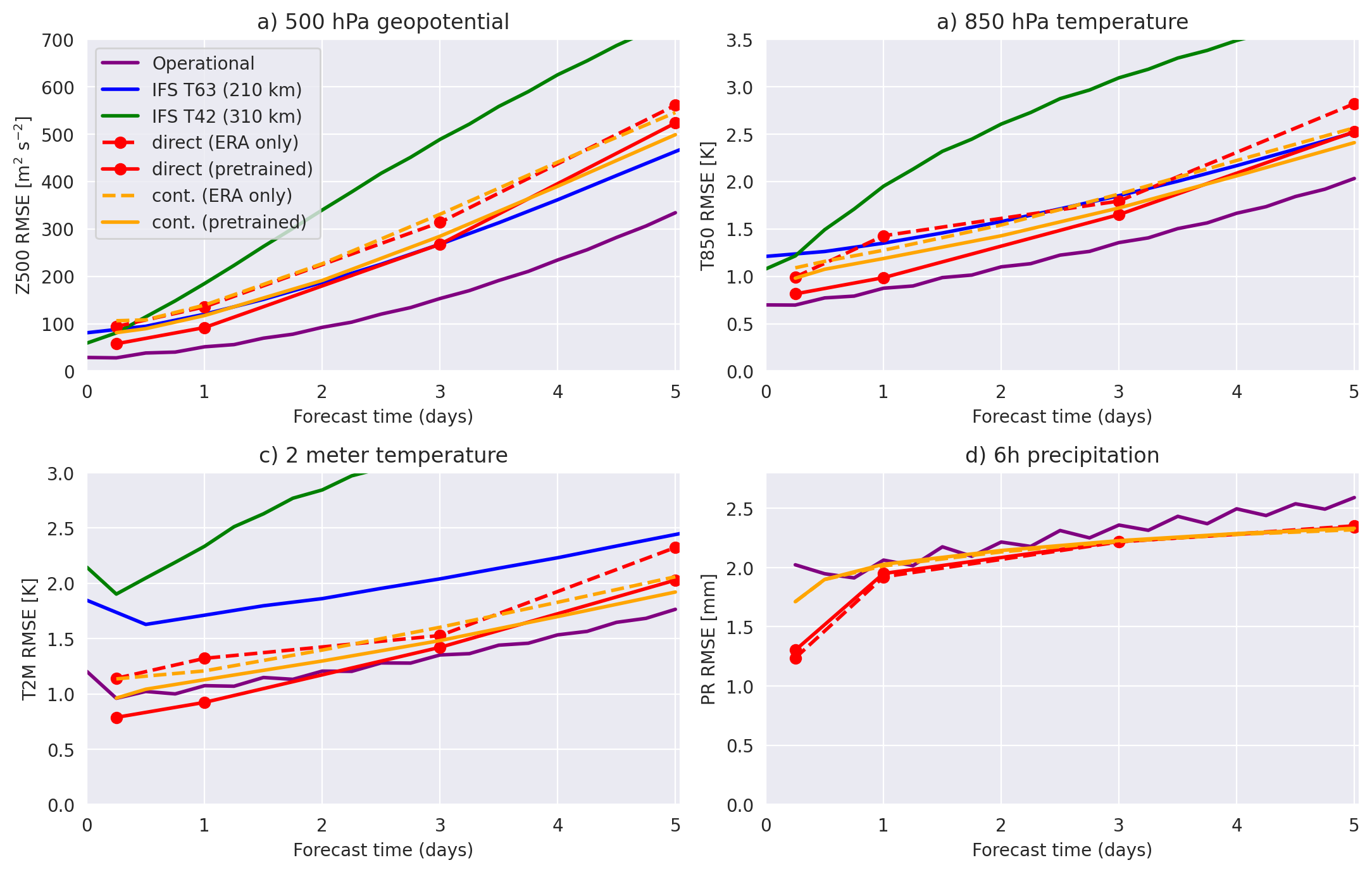}
\caption{Root mean squared error (RMSE) for a) Z500, b) T850, c) T2M and d) PR evaluated against ERA5 data.}
\label{fig:scores}
\end{figure}

\begin{figure}[bh!]
 \includegraphics[width=\linewidth]{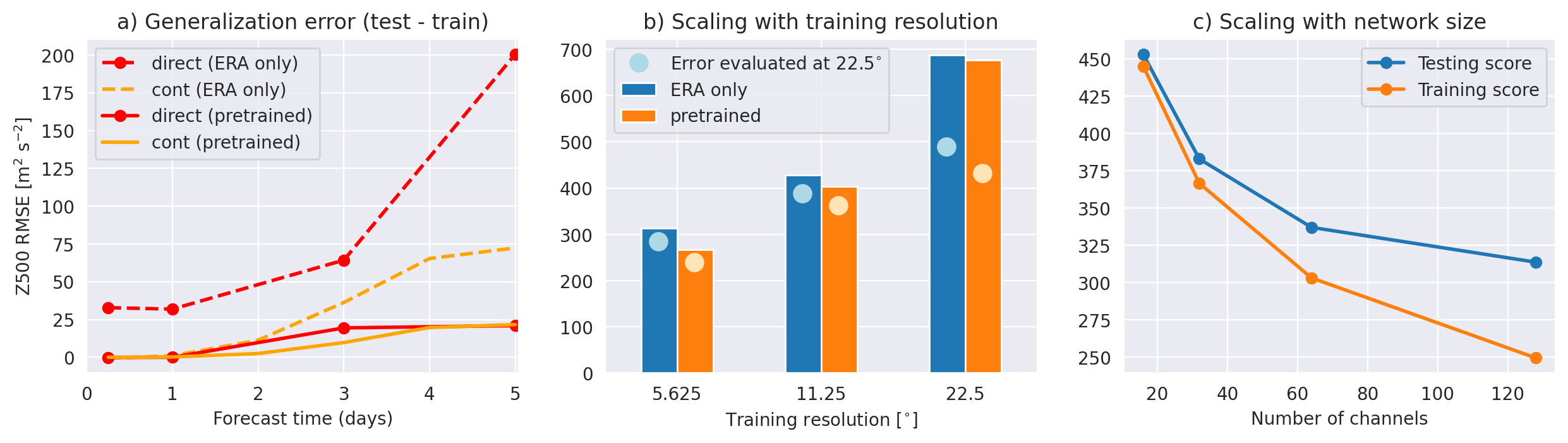}
\caption{a) Generalization error (testing minus training RMSE for Z500) b) RMSE of Z500 for networks trained with different resolution data. Bars show the RMSE computed at 5.625\degr resolution. For this the predictions from the lower resolution networks were upscaled. Dots show the RMSE evaluated at 22.5\degr for which all predictions were downscaled. c) RMSE of Z500 for different network architectures. y-axis has the same units (Z500 RMSE in m$^2$ s$^{-2}$) for all three panels.}
\label{fig:sensitivity}
\end{figure}

Figure~\ref{fig:scores} and Table~\ref{tab:rmse} show the results of the two networks on the WeatherBench metrics (ACC results can be found in Appendix Table~\ref{tab:acc}. Notably pretraining with CMIP data improves skill significantly for Z500, T850 and T2M over just using ERA data, with increasing impact for longer lead times. This is because overfitting, as measured by the difference between training and testing scores, tends to be worse for longer lead times (Fig.~\ref{fig:sensitivity}a). As the atmosphere becomes more chaotic for longer forecast horizons, similar initial conditions can lead to a wider range of outcomes. In the face of such uncertainty, a model that is trained to minimize the mean squared error, will tend to predict the mean of the distribution of possible outcomes. Our hypothesis is that for a wider distribution (longer forecast time) more training data is required to estimate the mean. In other words, if, because of the intrinsic unpredictability of the atmosphere, a broader range of outcome is physically plausible, then overfitting to individual outcomes encountered in the training data will lead to more overfitting than it would for shorter forecast times, where the plausible forecasts are closer together. Pretraining with climate model data helps to prevent overfitting and leads to better testing scores. Strikingly, even without fine-tuning on ERA data ("CMIP only" in Table~\ref{tab:rmse}) the testing scores computed on reanalysis data are not much or not at all worse than the "ERA only" networks. This shows that climate models, even though they do not exactly represent the real atmosphere, provide a good proxy for the general circulation of the atmosphere. For precipitation, pretraining does not improve skill. This is most likely because precipitation skill is low anyway and climate models might not represent precipitation as realistically as the large-scale circulation. Finally, it is important to note that RMSE and ACC are sub-optimal metrics for precipitation. 

Comparing the direct and continuous models, direct models tend to be better up to around 3 days forecast time, while the continuous models have more skill for longer forecast horizons. This difference also seems to be caused by overfitting. The continuous models without pretraining have a lower generalization error (Fig.~\ref{fig:sensitivity}a). One hypothesis as to why this is, could be that the fact that, in the continuous approach, a single model has to learn to make predictions for all forecast times acts as data-augmentation. Another plausible hypothesis is that the continuous networks also learn a time-evolution of the flow which helps regularize the network. With pretraining, the difference in the generalization error does not appear as large but this is potentially an artifact of using early-stopping for fine-tuning rather than a sign that the direct models do not overfit more. The two approaches, direct and continuous, therefore, represent a trade-off between specificity and generalization. One advantage of the continuous method is that arbitrary forecast times within the training range can be chosen. However, using the continuous network to predict beyond its training range quickly leads to large errors (not shown).

The models presented in this paper outperform the simple machine learning baselines from the original WeatherBench paper and also the approach of \citeA{Weyn2020} (Table~\ref{tab:rmse}). However, as mentioned previously, their model is an iterative model. Technically this is a more difficult approach because training a neural network for shirt-term predictions (in their case 6\,h) and then calling it iterativly can lead to self-amplifying errors. In fact, this is what we observed when trying this with our model architecture. Further, this requires the output vector to match the input vector, greatly increasing the number of variables to predict which can lead to a loss in specificity. \citeA{Weyn2020} trained the model over two time steps. This, however, quickly becomes very computationally expensive for large model such as the ones in this paper. Therefore, if the goal is simply to predict a certain field at a predefined time ahead, the direct and continuous approaches will likely lead to better results. On the other hand, iterative models can be used to make arbitrarily long predictions which opens up a range of potential use cases.

\begin{figure}[th!]
 \includegraphics[width=\linewidth]{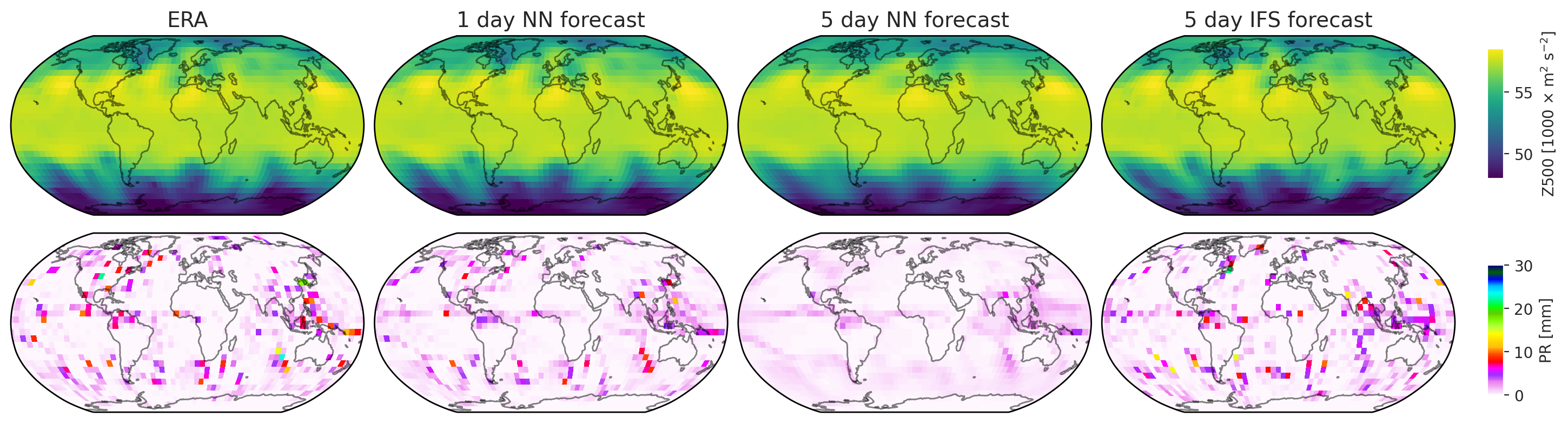}
\caption{Sample forecasts valid at 1 July 2018 00UTC for 500\,hPa geopotential (top row) and 6\,h accumulated precipitation (bottom row). 1 and 5 day pretrained, direct neural network forecasts are compared to the 5 day operational IFS forecast and the ERA5 ground truth.}
\label{fig:examples}
\end{figure}

Finally, it is interesting to discuss the performance of our models compared to the physical baselines. For Z500 and T850, the skill is comparable to the T63 model. For T2M and PR, the skill is closer to the operational model baseline. However, there are big caveats to consider in this comparison. First, the physical models (operational IFS and T63) are initialized from slightly different initial conditions, leading to a non-zero error at $t=0$. In addition, the coarse resolution models T42 and T63 suffer from errors due to the conversion to spherical coordinates at coarse resolutions. Since error growth is initially exponential, this initial condition difference primarily affects short forecast times up to two days \cite{Zhang2007}. A likely more important consideration is that the T42 and T63 models were not tuned for this resolution. This is in contrast to the operational IFS model which is carefully tuned over many years. This means that tuning the lower resolution IFS models would almost certainly lead to increased skill, however it is hard to estimate how much. On the other hand, our models are trained at significantly coarser resolutions and further hyper-parameter/architecture tuning would likely result in better scores. Another limitation is that statistical errors of the physical model were not removed by post-processing and that the evaluation was done at a very coarse grid. This is likely not so important for the upper-level variables Z500 and T850 but very important for surface variables like T2M and PR \cite{Hewson2020}. In data-driven forecasts the post-processing is implicitly performed. Lastly, it is important to consider that our models do not necessarily predict realistic fields. Fig.~\ref{fig:examples} one can see that with increasing lead time the predictions become more smoothed out and lose variability compared to the observations. This is another reflection of predicting the mean of the hypothetical forecast distribution as mentioned above. For geopotential and temperature this is especially grave in the extra-tropics where the largest natural fluctuations occur. For precipitation, a much more chaotic variable, the smoothing out is much more pronounced across the globe. At 5 days forecast time the data-driven forecast has no extreme events left. 

\section{Sensitivity to resolution and network size}
\label{sec:sensitivity}
It is interesting to ask how the results might change if the resolution was increased or larger networks were trained. Doing so, however, is technically very challenging and are outside the scope of this paper. We can however, assess the scaling to resolution and network size by using lower resolution and smaller networks. For this purpose we trained 3-day direct networks using 11.25\degr and 22.5\degr data but an otherwise identical training procedure (Fig.~\ref{fig:sensitivity}b). The skill drops with coarser resolution. This trend is present regardless whether the evaluation was done at 5.625\degr or 22.5\degr resolution with higher/lower resolution data interpolated to the evaluation resolution. This tendency makes sense since a higher data resolution provides better information to the network. One caveat of this sensitivity test is that we left the model architecture the same for these experiments, which means that the number of parameters relative to the size of the input/output vectors increases with coarser resolutions. 

To compare different network sizes we reduced the number of channels in each convolution from 128 to 64, 32 and 16 (Fig.~\ref{fig:sensitivity}c). The number of parameters decreases approximately by a factor 4 for each reduction. The testing skill increases with increasing network size but the trend flattens off and overfitting increases. This suggests that, while further improvements are certainly possible, there likely is a ceiling in skill for a given amount of training data. Note that the regularization parameters (weight decay and dropout) are the same across all network sizes. Another way to change the network size would be to change the number of layers. These experiments led to qualitatively similar results. Recent findings in deep learning \cite{Nakkiran2019} suggest that, further increasing network size can lead to lower testing losses despite increased overfitting. It would be interesting to see whether similar trends hold for this dataset.

\section{Interpretability}
\label{sec:interpretability}
The data-driven weather models predict weather with reasonable skill. One interesting question to ask is whether they do this for the "right reasons". To find out, we test which variables and which geographical region are important for the network to make a prediction. We do this by computing saliency maps \cite{Simonyan2013}. That is for each sample, we chose a point in space and a specific variable $p$, e.g. T850 over London. We then compute the gradient $G$ of this scalar $p$ with respect to the entire input array $X \in \mathcal{R}^{\mathrm{samples} \times \mathrm{lat} \times \mathrm{lon} \times \mathrm{variables}}$: $G = \partial p  / \partial X$ with the same shape as $X$. We do this analysis for two climatologically different locations: London, which is in the mid-latitudes and therefore influenced by eastwards-propagating Rossby waves and Barbados, located in the sub-tropical trade wind zones. This is done for different lead times using the pretrained direct networks.

It is important to highlight that the saliency method does not evaluate which inputs were most important for the prediction but rather which changes in the input would most affect the output. For a discussion on the differences, see \cite{Ebert-Uphoff2020}. For the purposes of this paper, the saliency method is appropriate since it allows us to evaluate effect of small input perturbations which is closely related to the body of work on adjoint sensitivity \cite{Ancell2007}.

\begin{figure}[th!]
 \includegraphics[width=\linewidth]{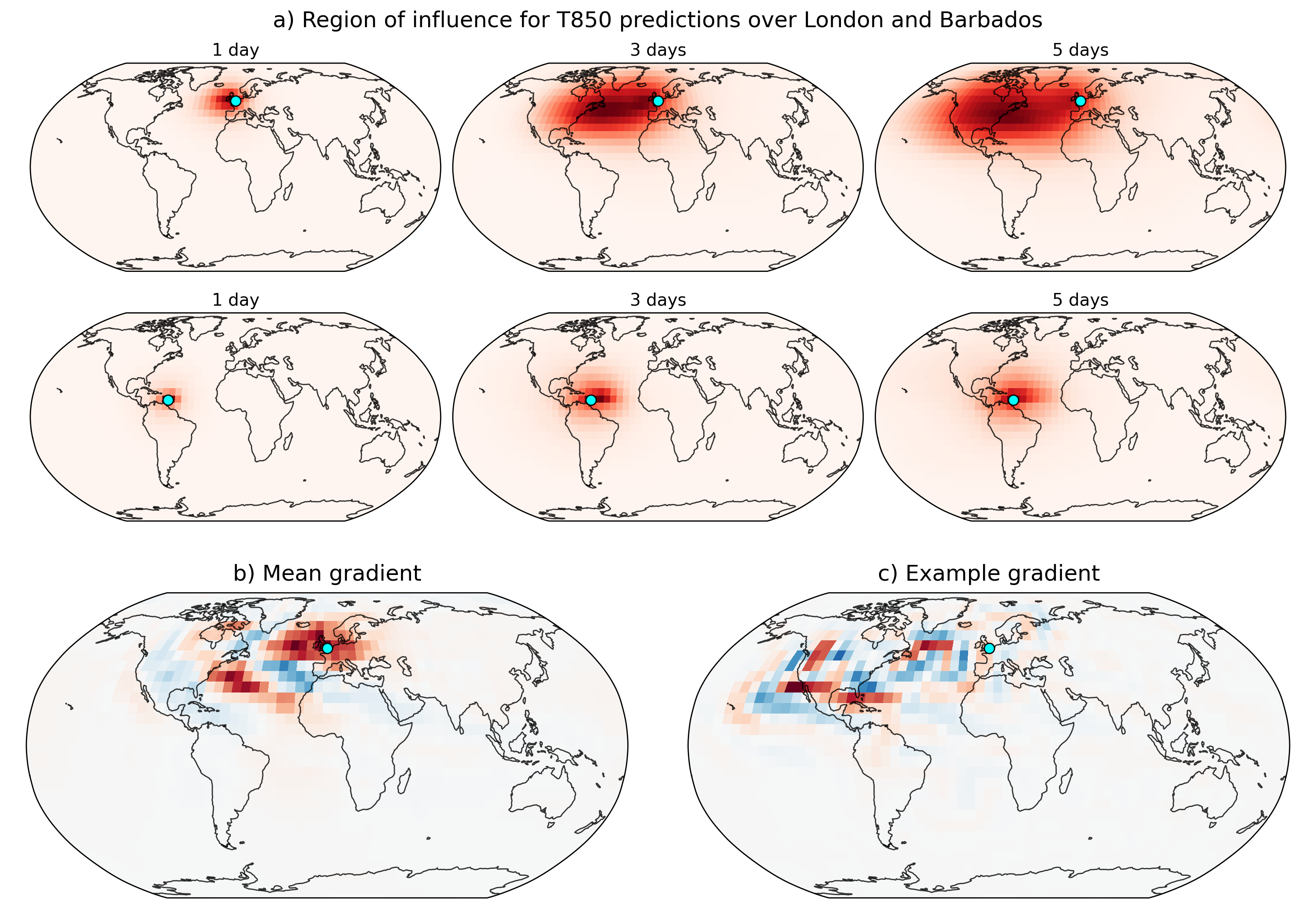}
\caption{Saliency plots: a) The region of influence $\overline{|G|}$ (see text for explanation) of T850 over London and Barbados with respect to all input variables (averaged). b) Mean gradient over time $\bar{G}$ of T850 with respect to Z500 over London. c) Sample gradient $G$ of Z500 with respect to 250\,hPa geopotential for 8 January 2017 12:00UTC.}
\label{fig:saliency}
\end{figure}

\begin{figure}[th!]
 \includegraphics[width=\linewidth]{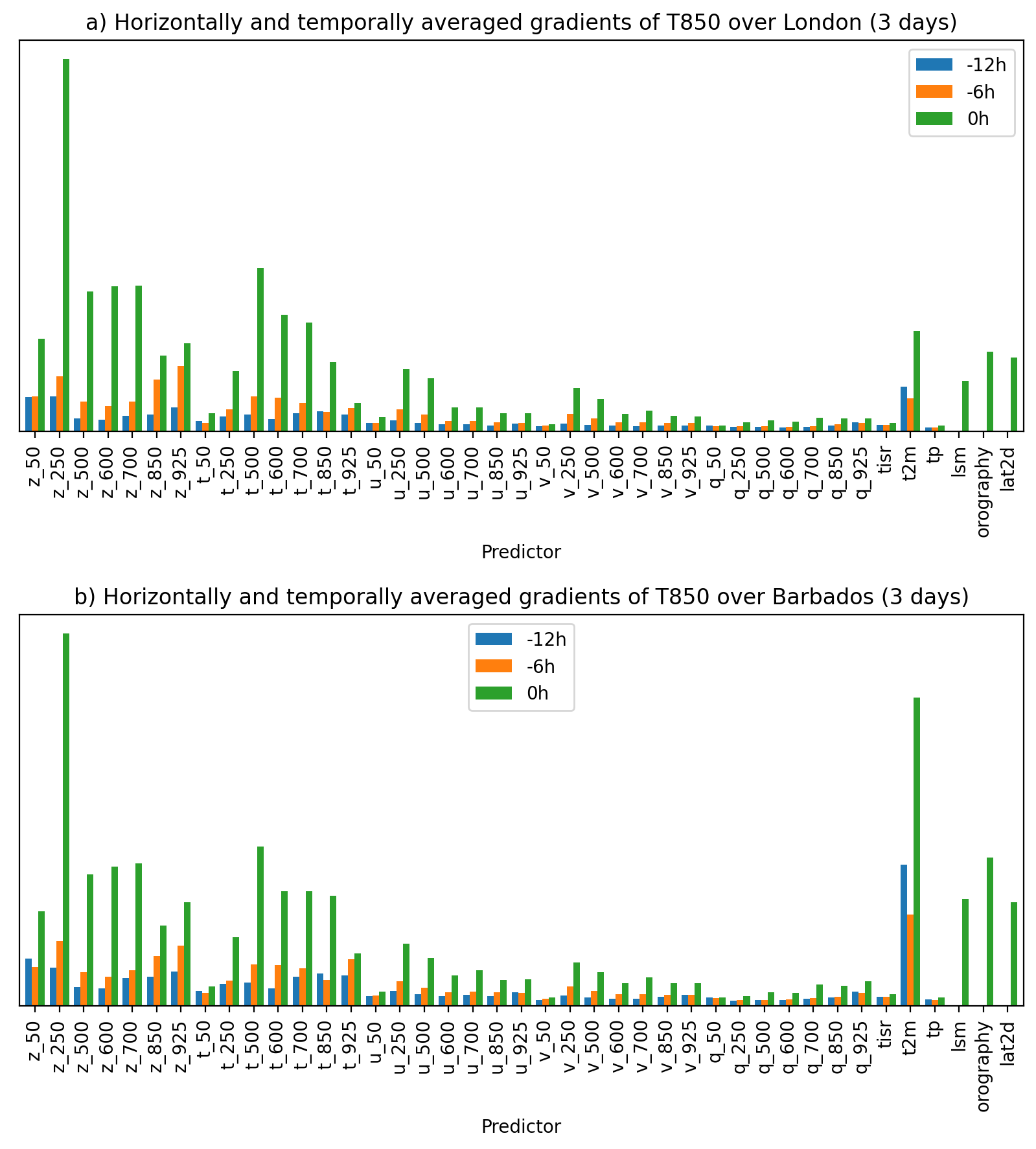}
\caption{Horizontally averaged saliency $\overline{|G|}$ of T850 over a) London b) Barbados for 3 day direct forecasts. tisr is the top-of-atmosphere incoming solar radiation, tp is precipitation, lsm is the land sea mask.}
\label{fig:saliency_bar}
\end{figure}

First, we investigate the region of influence by computing the mean absolute gradient of T850 over all samples $\overline{|G|} = 1/N_{\mathrm{samples}} \sum_i |G_i|$ and then taking the mean over all input variables (Fig.~\ref{fig:saliency}a). Because we compute the gradients for the normalized inputs, the different variables and levels should be comparable in scale and the gradients are dimensionless. It is important to highlight that the saliency analysis is primarily of qualitative nature. The resulting maps show that the networks tends to look at physically reasonable geographical regions. For London the region of influence extends towards the West with increasing forecast time. This is in line with our physical understanding of eastwards traveling Rossby waves being a key factor for weather in the mid-latitudes. Further we can look at the mean gradient $\bar{G} = 1/N_{\mathrm{samples}} \sum_i G_i$ of a specific input variable, in this case Z500, for 3 days forecast time (Fig.~\ref{fig:saliency}b). Here we see a positive-negative pattern across the Atlantic. Physically, one could interpret this as the signature of Rossby phase shifts influencing the temperature over London several days ahead. Over Barbados the region of influence looks very different and more circular which is in accordance with calmer meteorological conditions in the subtropics without a persistent preferred wind direction. Note that $\overline{|G|}$ is a mean over all seasons so that seasonally prevailing regimes do not show up.

We can also take the horizontal mean of $\bar{|G|}$ to obtain the mean influence of each normalized input variable (Fig.~\ref{fig:saliency_bar}). Geopotential and temperature show the largest gradients on average. Specifically changes in the geopotential at 250\,hPa appear to have a large effect. This is reasonable since 250\,hPa is close to the tropopause and changes in the tropopause height are known to be influential for medium-range weather evolution \cite{Hoskins1985}. Further, the gradient analysis shows that T2M is important for Barbados which reflects the importance of the ocean temperatures. Comparing, the influence of the inputs at the current time step $t$, $t-6$h and $t-12$h, the current time step is much more important than earlier time steps. The confirms our empirical findings that adding these previous time steps only improved the scores marginally (not shown). 

So far, all results are in agreement with physical reasoning and similar results could be expected to come out of adjoint sensitivity studies with physical models. However, looking at $G$ for individual samples, it is evident that this is not always the case. Fig.~\ref{fig:saliency}c shows the gradient of T850 over London with respect the 250\,hPa geopotential for a 3 day forecast. Significant gradients stretch across the Atlantic and North America all the way to Hawaii. This extent of information propagation within 3 days is rather unphysical. Studies using physical models typically estimate that it takes perturbations 5--6 days to cross the Atlantic \cite{Rodwell2013}. These results suggest that while the network, on average, learns physically plausible connections from data it appears to make unphysical connections for some samples. This makes sense since, in our setup, the network purely learns correlations between input and output images and there is nothing stopping it from learning "unphysical" correlations. If, for example, a certain pattern over eastern North America - which likely has an influence on European weather 3 days layer - also concurs in the training data with some pattern over the eastern Pacific, the network will pick up that connection between Pacific and European weather even if it might not be a causal relationship. In a way, such "unphysical" relations are also a sign of overfitting.

\section{Discussion and Conclusion}
\label{sec:conclusion}
In this paper, we presented a data-driven method for medium-range weather forecasting using a Resnet neural network architecture. Specifically, we trained models to predict 500\,hPa geopotential, 850\,hPa temperature, 2\,m temperature and precipitation up to 5 days ahead following the WeatherBench challenge. To avoid overfitting we pretrained the networks using climate model data from the CMIP archive. Our models set a new data-driven state-of-the-art for WeatherBench. Most previous approaches on similar problems used a U-Net \cite{Ronneberger2015} architecture, which in our experiments did not work as well as a simple Resnet without any changes in dimensionality. Compared to physical models, the Resnet achieves comparable scores to a physical model at comparable resolution. However, it is important not to over-interpret these results for several reasons discussed in the paper. More detailed evaluation would also be needed to accurately compare the two. The focus in this paper is primarily on the challenge set by WeatherBench.

It is more interesting to discuss the relevance of the findings presented here for data-driven weather forecasting in the future. It appears that with sufficient training data for pretraining purely data-driven forecasting can achieve reasonable skill. Our scaling analysis indicates that going to higher resolutions and larger networks leads to better scores. It is an interesting question whether the resolution scaling continues for higher resolutions than those considered here. However, the increased overfitting for larger networks already suggests that large amounts of data are required to train competitive data-driven models. One can also assume that larger models are needed for higher-resolutions to maintain a reasonable receptive field. Here, we used climate model simulations to combat overfitting. Current CMIP models, however, are run at around 100\,km resolution, and therefore cannot be used for forecasts at higher-resolutions. There are several atmosphere-only climate simulations \cite{Haarsma2016} run at resolutions comparable to the ERA5 resolution of 25\,km. It can be assumed that using all this available data at the highest possible resolution for training would greatly increase the forecast skill of data-driven methods. However, for the resolutions of current operational NWP models (10\,km) is it unlikely that there is sufficient data to challenge these models (see \citeA{Palmer2020} for a theoretical argument). As an aside, even if data-driven models matched physical models at forecasting, creating an initial condition currently requires data-assimilation systems that are currently based on physical models.

However, the findings regarding relative skill of data-driven versus physical forecasting are specific to the particular problem at hand. Data-driven methods could still play a large role in the broad field of weather forecasting. Two crucial questions to ask are how much training data is available for a particular problem and how much potential there is to improve upon physical approaches. For medium-range forecasting, physical modeling has achieved impressive skill in recent decades which makes it hard to do better with the observations at hand. Other task in numerical weather prediction could offer a much bigger potential for data-driven methods.

\section*{Open data}
The dataset is available at \url{https://mediatum.ub.tum.de/1524895} \cite{Rasp2020Data}. The code for the WeatherBench challenge is at \url{https://github.com/pangeo-data/WeatherBench}. The code for this paper specifically is at \url{https://github.com/raspstephan/WeatherBench}.

\appendix
\section{ACC results}

Table~\ref{tab:acc} shows the ACC skill for all experiments. ACC is defined as 
\begin{equation}
    \mathrm{ACC} = \frac{\sum_{i, j, k} L(j)  f'_{i, j, k}  t'_{i, j, k}}{\sqrt{\sum_{i, j, k} L(j)  f'^{2}_{i, j, k} \sum_{i, j, k} L(j)  t'^{2}_{i, j, k}}}
\end{equation}
where the prime $'$ denotes the difference to the climatology. Here the climatology is defined as $\mathrm{climatology}_{j, k} = \frac{1}{N_{time}} \sum t_{j, k}$. 

\begin{table}[th]
\caption{ACC for 3 and 5 days forecast time. All forecasts evaluated at 5.625\degr resolution.}
\begin{tabular}{l||l|l|l|l}
 & \multicolumn{4}{c}{\hspace{0pt} Latitude-weighted ACC (3 days / 5 days)} \\
\textbf{Model} & \textbf{Z500 [m$^2$ s$^{-2}$]} & \textbf{T850 [K]} & \textbf{T2M [K]} & \textbf{PR [mm]} \\ \hline \hline
Persistence & 0.62 / 0.53 & 0.69 / 0.65 & 0.88 / 0.85 & 0.06/0.06 \\ \hline
Climatology & 0 & 0  & 0 & 0\\ \hline
Weekly climatology & 0.65 & 0.77  & 0.85 & 0.16 \\ \hline

IFS T42 & 0.90 / 0.78 & 0.86 / 0.78  & 0.87 / 0.83 & \\ \hline
IFS T63 & 0.97 / 0.91 & 0.94 / 0.90  & 0.94 / 0.92 & \\ \hline
Operational IFS & \textbf{0.99} / \textbf{0.95} & \textbf{0.97} / \textbf{0.93} & \textbf{0.98} / \textbf{0.96} & \textbf{0.43} / \textbf{0.30} \\ \hline \hline
Direct (ERA only) & 0.96 / 0.85 & 0.94 / 0.86 & 0.97 / 0.92 & \textbf{0.55} / 0.24 \\ \hline
Direct (CMIP only) & 0.95 / 0.85 & 0.93 / 0.86 & 0.95 / 0.92 & 0.32 / 0.20 \\ \hline
Direct (pretrained) & \textbf{0.97} / 0.87 & \textbf{0.95} / 0.89 & \textbf{0.97} / 0.94 & 0.45 / \textbf{0.29} \\ \hline
Continuous (ERA only) & 0.95 / 0.86 & 0.94 / 0.88 & 0.96 / 0.94 & 0.41 / 0.29 \\ \hline
Continuous (CMIP only) & 0.95 / 0.86 & 0.93 / 0.87 & 0.93 / 0.91 & 0.41 / 0.29 \\ \hline
Continuous (pretrained) & 0.96 / \textbf{0.88} & \textbf{0.95} / \textbf{0.90} & 0.97 / \textbf{0.95} & \textbf{0.41} / 0.28 \\ \hline

\end{tabular}
\label{tab:acc}
\end{table}

\acknowledgments
We would like to thank Sebastian Scher and David Greenberg for their valuable comments on the paper as well as George Craig for discussing the saliency analysis. Stephan Rasp acknowledges funding from the German Research Foundation (DFG) under grant no. 426852073. The dataset is available at \url{https://mediatum.ub.tum.de/1524895} \cite{Rasp2020Data}. The code for the WeatherBench challenge is at \url{https://github.com/pangeo-data/WeatherBench}. The code for this paper specifically is at \url{https://github.com/raspstephan/WeatherBench}.


%
%

\bibliography{references.bib}

\begin{thebibliography}{}

\bibitem [\protect \citeauthoryear {%
Ancell%
\ \BBA {} Hakim%
}{%
Ancell%
\ \BBA {} Hakim%
}{%
{\protect \APACyear {2007}}%
}]{%
Ancell2007}
\APACinsertmetastar {%
Ancell2007}%
\begin{APACrefauthors}%
Ancell, B.%
\BCBT {}\ \BBA {} Hakim, G\BPBI J.%
\end{APACrefauthors}%
\unskip\
\newblock
\APACrefYearMonthDay{2007}{}{}.
\newblock
{\BBOQ}\APACrefatitle {{Comparing Adjoint- and Ensemble-Sensitivity Analysis
  with Applications to Observation Targeting}} {{Comparing Adjoint- and
  Ensemble-Sensitivity Analysis with Applications to Observation
  Targeting}}.{\BBCQ}
\newblock
\APACjournalVolNumPages{Monthly Weather Review}{135}{}{4117--4134}.
\newblock
\begin{APACrefDOI} \doi{10.1175/2007MWR1904.1} \end{APACrefDOI}
\PrintBackRefs{\CurrentBib}

\bibitem [\protect \citeauthoryear {%
Bauer%
, Thorpe%
\BCBL {}\ \BBA {} Brunet%
}{%
Bauer%
\ \protect \BOthers {.}}{%
{\protect \APACyear {2015}}%
}]{%
Bauer2015a}
\APACinsertmetastar {%
Bauer2015a}%
\begin{APACrefauthors}%
Bauer, P.%
, Thorpe, A.%
\BCBL {}\ \BBA {} Brunet, G.%
\end{APACrefauthors}%
\unskip\
\newblock
\APACrefYearMonthDay{2015}{9}{}.
\newblock
{\BBOQ}\APACrefatitle {{The quiet revolution of numerical weather prediction}}
  {{The quiet revolution of numerical weather prediction}}.{\BBCQ}
\newblock
\APACjournalVolNumPages{Nature}{525}{7567}{47--55}.
\newblock
\begin{APACrefURL} \url{http://www.nature.com/doifinder/10.1038/nature14956}
  \end{APACrefURL}
\newblock
\begin{APACrefDOI} \doi{10.1038/nature14956} \end{APACrefDOI}
\PrintBackRefs{\CurrentBib}

\bibitem [\protect \citeauthoryear {%
Dueben%
\ \BBA {} Bauer%
}{%
Dueben%
\ \BBA {} Bauer%
}{%
{\protect \APACyear {2018}}%
}]{%
Dueben2018}
\APACinsertmetastar {%
Dueben2018}%
\begin{APACrefauthors}%
Dueben, P\BPBI D.%
\BCBT {}\ \BBA {} Bauer, P.%
\end{APACrefauthors}%
\unskip\
\newblock
\APACrefYearMonthDay{2018}{}{}.
\newblock
{\BBOQ}\APACrefatitle {{Challenges and design choices for global weather and
  climate models based on machine learning}} {{Challenges and design choices
  for global weather and climate models based on machine learning}}.{\BBCQ}
\newblock
\APACjournalVolNumPages{Geosci. Model Dev.}{}{}{}.
\newblock
\begin{APACrefURL}
  \url{https://www.geosci-model-dev-discuss.net/gmd-2018-148/gmd-2018-148.pdf}
  \end{APACrefURL}
\newblock
\begin{APACrefDOI} \doi{10.5194/gmd-2018-148} \end{APACrefDOI}
\PrintBackRefs{\CurrentBib}

\bibitem [\protect \citeauthoryear {%
Ebert-Uphoff%
\ \BBA {} Hilburn%
}{%
Ebert-Uphoff%
\ \BBA {} Hilburn%
}{%
{\protect \APACyear {2020}}%
}]{%
Ebert-Uphoff2020}
\APACinsertmetastar {%
Ebert-Uphoff2020}%
\begin{APACrefauthors}%
Ebert-Uphoff, I.%
\BCBT {}\ \BBA {} Hilburn, K\BPBI A.%
\end{APACrefauthors}%
\unskip\
\newblock
\APACrefYearMonthDay{2020}{5}{}.
\newblock
{\BBOQ}\APACrefatitle {{Evaluation, Tuning and Interpretation of Neural
  Networks for Meteorological Applications}} {{Evaluation, Tuning and
  Interpretation of Neural Networks for Meteorological Applications}}.{\BBCQ}
\newblock
\begin{APACrefURL} \url{http://arxiv.org/abs/2005.03126} \end{APACrefURL}
\PrintBackRefs{\CurrentBib}

\bibitem [\protect \citeauthoryear {%
Eyring%
\ \protect \BOthers {.}}{%
Eyring%
\ \protect \BOthers {.}}{%
{\protect \APACyear {2016}}%
}]{%
Eyring2016}
\APACinsertmetastar {%
Eyring2016}%
\begin{APACrefauthors}%
Eyring, V.%
, Bony, S.%
, Meehl, G\BPBI A.%
, Senior, C\BPBI A.%
, Stevens, B.%
, Stouffer, R\BPBI J.%
\BCBL {}\ \BBA {} Taylor, K\BPBI E.%
\end{APACrefauthors}%
\unskip\
\newblock
\APACrefYearMonthDay{2016}{5}{}.
\newblock
{\BBOQ}\APACrefatitle {{Overview of the Coupled Model Intercomparison Project
  Phase 6 (CMIP6) experimental design and organization}} {{Overview of the
  Coupled Model Intercomparison Project Phase 6 (CMIP6) experimental design and
  organization}}.{\BBCQ}
\newblock
\APACjournalVolNumPages{Geoscientific Model Development}{9}{5}{1937--1958}.
\newblock
\begin{APACrefURL} \url{https://www.geosci-model-dev.net/9/1937/2016/}
  \end{APACrefURL}
\newblock
\begin{APACrefDOI} \doi{10.5194/gmd-9-1937-2016} \end{APACrefDOI}
\PrintBackRefs{\CurrentBib}

\bibitem [\protect \citeauthoryear {%
Gr{\"{o}}nquist%
\ \protect \BOthers {.}}{%
Gr{\"{o}}nquist%
\ \protect \BOthers {.}}{%
{\protect \APACyear {2020}}%
}]{%
Gronquist2020}
\APACinsertmetastar {%
Gronquist2020}%
\begin{APACrefauthors}%
Gr{\"{o}}nquist, P.%
, Yao, C.%
, Ben-Nun, T.%
, Dryden, N.%
, Dueben, P.%
, Li, S.%
\BCBL {}\ \BBA {} Hoefler, T.%
\end{APACrefauthors}%
\unskip\
\newblock
\APACrefYearMonthDay{2020}{5}{}.
\newblock
{\BBOQ}\APACrefatitle {{Deep Learning for Post-Processing Ensemble Weather
  Forecasts}} {{Deep Learning for Post-Processing Ensemble Weather
  Forecasts}}.{\BBCQ}
\newblock
\begin{APACrefURL} \url{http://arxiv.org/abs/2005.08748} \end{APACrefURL}
\PrintBackRefs{\CurrentBib}

\bibitem [\protect \citeauthoryear {%
Haarsma%
\ \protect \BOthers {.}}{%
Haarsma%
\ \protect \BOthers {.}}{%
{\protect \APACyear {2016}}%
}]{%
Haarsma2016}
\APACinsertmetastar {%
Haarsma2016}%
\begin{APACrefauthors}%
Haarsma, R\BPBI J.%
, Roberts, M\BPBI J.%
, Vidale, P\BPBI L.%
, Senior, C\BPBI A.%
, Bellucci, A.%
, Bao, Q.%
\BDBL {}von Storch, J\BHBI S.%
\end{APACrefauthors}%
\unskip\
\newblock
\APACrefYearMonthDay{2016}{11}{}.
\newblock
{\BBOQ}\APACrefatitle {{High Resolution Model Intercomparison Project
  (HighResMIP v1.0) for CMIP6}} {{High Resolution Model Intercomparison
  Project (HighResMIP v1.0) for CMIP6}}.{\BBCQ}
\newblock
\APACjournalVolNumPages{Geoscientific Model Development}{9}{11}{4185--4208}.
\newblock
\begin{APACrefURL} \url{https://gmd.copernicus.org/articles/9/4185/2016/}
  \end{APACrefURL}
\newblock
\begin{APACrefDOI} \doi{10.5194/gmd-9-4185-2016} \end{APACrefDOI}
\PrintBackRefs{\CurrentBib}

\bibitem [\protect \citeauthoryear {%
He%
, Zhang%
, Ren%
\BCBL {}\ \BBA {} Sun%
}{%
He%
\ \protect \BOthers {.}}{%
{\protect \APACyear {2015}}%
}]{%
He2015}
\APACinsertmetastar {%
He2015}%
\begin{APACrefauthors}%
He, K.%
, Zhang, X.%
, Ren, S.%
\BCBL {}\ \BBA {} Sun, J.%
\end{APACrefauthors}%
\unskip\
\newblock
\APACrefYearMonthDay{2015}{12}{}.
\newblock
{\BBOQ}\APACrefatitle {{Deep Residual Learning for Image Recognition}} {{Deep
  Residual Learning for Image Recognition}}.{\BBCQ}
\newblock
\begin{APACrefURL} \url{http://arxiv.org/abs/1512.03385} \end{APACrefURL}
\PrintBackRefs{\CurrentBib}

\bibitem [\protect \citeauthoryear {%
Hersbach%
\ \protect \BOthers {.}}{%
Hersbach%
\ \protect \BOthers {.}}{%
{\protect \APACyear {2020}}%
}]{%
Hersbach2020}
\APACinsertmetastar {%
Hersbach2020}%
\begin{APACrefauthors}%
Hersbach, H.%
, Bell, B.%
, Berrisford, P.%
, Hirahara, S.%
, Hor{\'{a}}nyi, A.%
, Mu{\~{n}}oz‐Sabater, J.%
\BDBL {}Th{\'{e}}paut, J.%
\end{APACrefauthors}%
\unskip\
\newblock
\APACrefYearMonthDay{2020}{5}{}.
\newblock
{\BBOQ}\APACrefatitle {{The ERA5 Global Reanalysis}} {{The ERA5 Global
  Reanalysis}}.{\BBCQ}
\newblock
\APACjournalVolNumPages{Quarterly Journal of the Royal Meteorological
  Society}{}{}{qj.3803}.
\newblock
\begin{APACrefURL}
  \url{https://onlinelibrary.wiley.com/doi/abs/10.1002/qj.3803}
  \end{APACrefURL}
\newblock
\begin{APACrefDOI} \doi{10.1002/qj.3803} \end{APACrefDOI}
\PrintBackRefs{\CurrentBib}

\bibitem [\protect \citeauthoryear {%
Hewson%
\ \BBA {} Pillosu%
}{%
Hewson%
\ \BBA {} Pillosu%
}{%
{\protect \APACyear {2020}}%
}]{%
Hewson2020}
\APACinsertmetastar {%
Hewson2020}%
\begin{APACrefauthors}%
Hewson, T\BPBI D.%
\BCBT {}\ \BBA {} Pillosu, F\BPBI M.%
\end{APACrefauthors}%
\unskip\
\newblock
\APACrefYearMonthDay{2020}{3}{}.
\newblock
{\BBOQ}\APACrefatitle {{A new low-cost technique improves weather forecasts
  across the world}} {{A new low-cost technique improves weather forecasts
  across the world}}.{\BBCQ}
\newblock
\begin{APACrefURL} \url{http://arxiv.org/abs/2003.14397} \end{APACrefURL}
\PrintBackRefs{\CurrentBib}

\bibitem [\protect \citeauthoryear {%
Hoskins%
, McIntyre%
\BCBL {}\ \BBA {} Robertson%
}{%
Hoskins%
\ \protect \BOthers {.}}{%
{\protect \APACyear {1985}}%
}]{%
Hoskins1985}
\APACinsertmetastar {%
Hoskins1985}%
\begin{APACrefauthors}%
Hoskins, B\BPBI J.%
, McIntyre, M\BPBI E.%
\BCBL {}\ \BBA {} Robertson, A\BPBI W.%
\end{APACrefauthors}%
\unskip\
\newblock
\APACrefYearMonthDay{1985}{8}{}.
\newblock
{\BBOQ}\APACrefatitle {{On the use and significance of isentropic potential
  vorticity maps}} {{On the use and significance of isentropic potential
  vorticity maps}}.{\BBCQ}
\newblock
\APACjournalVolNumPages{Quarterly Journal of the Royal Meteorological
  Society}{111}{470}{877--946}.
\newblock
\begin{APACrefURL} \url{http://doi.wiley.com/10.1002/qj.49711147002}
  \end{APACrefURL}
\newblock
\begin{APACrefDOI} \doi{10.1002/qj.49711147002} \end{APACrefDOI}
\PrintBackRefs{\CurrentBib}

\bibitem [\protect \citeauthoryear {%
Hourdin%
\ \protect \BOthers {.}}{%
Hourdin%
\ \protect \BOthers {.}}{%
{\protect \APACyear {2017}}%
}]{%
Hourdin2017}
\APACinsertmetastar {%
Hourdin2017}%
\begin{APACrefauthors}%
Hourdin, F.%
, Mauritsen, T.%
, Gettelman, A.%
, Golaz, J\BHBI C.%
, Balaji, V.%
, Duan, Q.%
\BDBL {}Williamson, D.%
\end{APACrefauthors}%
\unskip\
\newblock
\APACrefYearMonthDay{2017}{3}{}.
\newblock
{\BBOQ}\APACrefatitle {{The Art and Science of Climate Model Tuning}} {{The Art
  and Science of Climate Model Tuning}}.{\BBCQ}
\newblock
\APACjournalVolNumPages{Bulletin of the American Meteorological
  Society}{98}{3}{589--602}.
\newblock
\begin{APACrefURL}
  \url{http://journals.ametsoc.org/doi/10.1175/BAMS-D-15-00135.1}
  \end{APACrefURL}
\newblock
\begin{APACrefDOI} \doi{10.1175/BAMS-D-15-00135.1} \end{APACrefDOI}
\PrintBackRefs{\CurrentBib}

\bibitem [\protect \citeauthoryear {%
Kalnay%
}{%
Kalnay%
}{%
{\protect \APACyear {2003}}%
}]{%
Kalnay2003}
\APACinsertmetastar {%
Kalnay2003}%
\begin{APACrefauthors}%
Kalnay, E.%
\end{APACrefauthors}%
\unskip\
\newblock
\APACrefYear{2003}.
\newblock
\APACrefbtitle {{Atmospheric modeling, data assimilation, and predictability}}
  {{Atmospheric modeling, data assimilation, and predictability}}\ (\BVOL~54).
\newblock
\begin{APACrefURL}
  \url{http://books.google.com/books?hl=en&amp;lr=&amp;id=Uqc7zC7NULMC&amp;oi=fnd&amp;pg=PR11&amp;dq=Atmospheric+modeling,+data+assimilation+and+predictability&amp;ots=lI5gpir1RV&amp;sig=FuhXqkYSMxhz2jLI2T8144HX6fs}
  \end{APACrefURL}
\PrintBackRefs{\CurrentBib}

\bibitem [\protect \citeauthoryear {%
Kingma%
\ \BBA {} Ba%
}{%
Kingma%
\ \BBA {} Ba%
}{%
{\protect \APACyear {2014}}%
}]{%
Kingma2014}
\APACinsertmetastar {%
Kingma2014}%
\begin{APACrefauthors}%
Kingma, D\BPBI P.%
\BCBT {}\ \BBA {} Ba, J.%
\end{APACrefauthors}%
\unskip\
\newblock
\APACrefYearMonthDay{2014}{12}{}.
\newblock
{\BBOQ}\APACrefatitle {{Adam: A Method for Stochastic Optimization}} {{Adam: A
  Method for Stochastic Optimization}}.{\BBCQ}
\newblock
\APACjournalVolNumPages{arXiv}{1412.6980}{}{}.
\newblock
\begin{APACrefURL} \url{http://arxiv.org/abs/1412.6980} \end{APACrefURL}
\PrintBackRefs{\CurrentBib}

\bibitem [\protect \citeauthoryear {%
McGovern%
\ \protect \BOthers {.}}{%
McGovern%
\ \protect \BOthers {.}}{%
{\protect \APACyear {2017}}%
}]{%
McGovern2017}
\APACinsertmetastar {%
McGovern2017}%
\begin{APACrefauthors}%
McGovern, A.%
, Elmore, K\BPBI L.%
, Gagne, D\BPBI J.%
, Haupt, S\BPBI E.%
, Karstens, C\BPBI D.%
, Lagerquist, R.%
\BDBL {}Williams, J\BPBI K.%
\end{APACrefauthors}%
\unskip\
\newblock
\APACrefYearMonthDay{2017}{10}{}.
\newblock
{\BBOQ}\APACrefatitle {{Using Artificial Intelligence to Improve Real-Time
  Decision-Making for High-Impact Weather}} {{Using Artificial Intelligence to
  Improve Real-Time Decision-Making for High-Impact Weather}}.{\BBCQ}
\newblock
\APACjournalVolNumPages{Bulletin of the American Meteorological
  Society}{98}{10}{2073--2090}.
\newblock
\begin{APACrefURL}
  \url{http://journals.ametsoc.org/doi/10.1175/BAMS-D-16-0123.1}
  \end{APACrefURL}
\newblock
\begin{APACrefDOI} \doi{10.1175/BAMS-D-16-0123.1} \end{APACrefDOI}
\PrintBackRefs{\CurrentBib}

\bibitem [\protect \citeauthoryear {%
Nakkiran%
\ \protect \BOthers {.}}{%
Nakkiran%
\ \protect \BOthers {.}}{%
{\protect \APACyear {2019}}%
}]{%
Nakkiran2019}
\APACinsertmetastar {%
Nakkiran2019}%
\begin{APACrefauthors}%
Nakkiran, P.%
, Kaplun, G.%
, Bansal, Y.%
, Yang, T.%
, Barak, B.%
\BCBL {}\ \BBA {} Sutskever, I.%
\end{APACrefauthors}%
\unskip\
\newblock
\APACrefYearMonthDay{2019}{12}{}.
\newblock
{\BBOQ}\APACrefatitle {{Deep Double Descent: Where Bigger Models and More Data
  Hurt}} {{Deep Double Descent: Where Bigger Models and More Data
  Hurt}}.{\BBCQ}
\newblock
\begin{APACrefURL} \url{http://arxiv.org/abs/1912.02292} \end{APACrefURL}
\PrintBackRefs{\CurrentBib}

\bibitem [\protect \citeauthoryear {%
Palmer%
}{%
Palmer%
}{%
{\protect \APACyear {2020}}%
}]{%
Palmer2020}
\APACinsertmetastar {%
Palmer2020}%
\begin{APACrefauthors}%
Palmer, T.%
\end{APACrefauthors}%
\unskip\
\newblock
\APACrefYearMonthDay{2020}{7}{}.
\newblock
{\BBOQ}\APACrefatitle {{A Vision for Numerical Weather Prediction in 2030}} {{A
  Vision for Numerical Weather Prediction in 2030}}.{\BBCQ}
\newblock
\begin{APACrefURL} \url{http://arxiv.org/abs/2007.04830} \end{APACrefURL}
\PrintBackRefs{\CurrentBib}

\bibitem [\protect \citeauthoryear {%
Rasp%
\ \protect \BOthers {.}}{%
Rasp%
\ \protect \BOthers {.}}{%
{\protect \APACyear {2020}}%
{\protect \APACexlab {{\protect \BCnt {1}}}}}]{%
Rasp2020Data}
\APACinsertmetastar {%
Rasp2020Data}%
\begin{APACrefauthors}%
Rasp, S.%
, Dueben, P\BPBI D.%
, Scher, S.%
, Weyn, J\BPBI A.%
, Mouatadid, S.%
\BCBL {}\ \BBA {} Thuerey, N.%
\end{APACrefauthors}%
\unskip\
\newblock
\APACrefYearMonthDay{2020{\protect \BCnt {1}}}{}{}.
\newblock
\APACrefbtitle {{WeatherBench: A benchmark dataset for data-driven weather
  forecasting}.} {{WeatherBench: A benchmark dataset for data-driven weather
  forecasting}.}
\newblock
\APACaddressPublisher{}{Technical University of Munich}.
\newblock
\begin{APACrefURL} \url{https://mediatum.ub.tum.de/1524895} \end{APACrefURL}
\newblock
\begin{APACrefDOI} \doi{10.14459/2019mp1524895} \end{APACrefDOI}
\PrintBackRefs{\CurrentBib}

\bibitem [\protect \citeauthoryear {%
Rasp%
\ \protect \BOthers {.}}{%
Rasp%
\ \protect \BOthers {.}}{%
{\protect \APACyear {2020}}%
{\protect \APACexlab {{\protect \BCnt {2}}}}}]{%
Rasp2020a}
\APACinsertmetastar {%
Rasp2020a}%
\begin{APACrefauthors}%
Rasp, S.%
, Dueben, P\BPBI D.%
, Scher, S.%
, Weyn, J\BPBI A.%
, Mouatadid, S.%
\BCBL {}\ \BBA {} Thuerey, N.%
\end{APACrefauthors}%
\unskip\
\newblock
\APACrefYearMonthDay{2020{\protect \BCnt {2}}}{2}{}.
\newblock
{\BBOQ}\APACrefatitle {{WeatherBench: A benchmark dataset for data-driven
  weather forecasting}} {{WeatherBench: A benchmark dataset for data-driven
  weather forecasting}}.{\BBCQ}
\newblock
\begin{APACrefURL} \url{http://arxiv.org/abs/2002.00469} \end{APACrefURL}
\PrintBackRefs{\CurrentBib}

\bibitem [\protect \citeauthoryear {%
Rasp%
\ \BBA {} Lerch%
}{%
Rasp%
\ \BBA {} Lerch%
}{%
{\protect \APACyear {2018}}%
}]{%
Rasp2018d}
\APACinsertmetastar {%
Rasp2018d}%
\begin{APACrefauthors}%
Rasp, S.%
\BCBT {}\ \BBA {} Lerch, S.%
\end{APACrefauthors}%
\unskip\
\newblock
\APACrefYearMonthDay{2018}{11}{}.
\newblock
{\BBOQ}\APACrefatitle {{Neural Networks for Postprocessing Ensemble Weather
  Forecasts}} {{Neural Networks for Postprocessing Ensemble Weather
  Forecasts}}.{\BBCQ}
\newblock
\APACjournalVolNumPages{Monthly Weather Review}{146}{11}{3885--3900}.
\newblock
\begin{APACrefURL}
  \url{http://journals.ametsoc.org/doi/10.1175/MWR-D-18-0187.1}
  \end{APACrefURL}
\newblock
\begin{APACrefDOI} \doi{10.1175/MWR-D-18-0187.1} \end{APACrefDOI}
\PrintBackRefs{\CurrentBib}

\bibitem [\protect \citeauthoryear {%
Rodwell%
\ \protect \BOthers {.}}{%
Rodwell%
\ \protect \BOthers {.}}{%
{\protect \APACyear {2013}}%
}]{%
Rodwell2013}
\APACinsertmetastar {%
Rodwell2013}%
\begin{APACrefauthors}%
Rodwell, M\BPBI J.%
, Magnusson, L.%
, Bauer, P.%
, Bechtold, P.%
, Bonavita, M.%
, Cardinali, C.%
\BDBL {}Wedi, N.%
\end{APACrefauthors}%
\unskip\
\newblock
\APACrefYearMonthDay{2013}{9}{}.
\newblock
{\BBOQ}\APACrefatitle {{Characteristics of Occasional Poor Medium-Range Weather
  Forecasts for Europe}} {{Characteristics of Occasional Poor Medium-Range
  Weather Forecasts for Europe}}.{\BBCQ}
\newblock
\APACjournalVolNumPages{Bulletin of the American Meteorological
  Society}{94}{9}{1393--1405}.
\newblock
\begin{APACrefURL}
  \url{http://journals.ametsoc.org/doi/abs/10.1175/BAMS-D-12-00099.1}
  \end{APACrefURL}
\newblock
\begin{APACrefDOI} \doi{10.1175/BAMS-D-12-00099.1} \end{APACrefDOI}
\PrintBackRefs{\CurrentBib}

\bibitem [\protect \citeauthoryear {%
Ronneberger%
, Fischer%
\BCBL {}\ \BBA {} Brox%
}{%
Ronneberger%
\ \protect \BOthers {.}}{%
{\protect \APACyear {2015}}%
}]{%
Ronneberger2015}
\APACinsertmetastar {%
Ronneberger2015}%
\begin{APACrefauthors}%
Ronneberger, O.%
, Fischer, P.%
\BCBL {}\ \BBA {} Brox, T.%
\end{APACrefauthors}%
\unskip\
\newblock
\APACrefYearMonthDay{2015}{5}{}.
\newblock
{\BBOQ}\APACrefatitle {{U-Net: Convolutional Networks for Biomedical Image
  Segmentation}} {{U-Net: Convolutional Networks for Biomedical Image
  Segmentation}}.{\BBCQ}
\newblock
\begin{APACrefURL} \url{http://arxiv.org/abs/1505.04597} \end{APACrefURL}
\PrintBackRefs{\CurrentBib}

\bibitem [\protect \citeauthoryear {%
Scher%
}{%
Scher%
}{%
{\protect \APACyear {2018}}%
}]{%
Scher2018}
\APACinsertmetastar {%
Scher2018}%
\begin{APACrefauthors}%
Scher, S.%
\end{APACrefauthors}%
\unskip\
\newblock
\APACrefYearMonthDay{2018}{11}{}.
\newblock
{\BBOQ}\APACrefatitle {{Toward Data‐Driven Weather and Climate Forecasting:
  Approximating a Simple General Circulation Model With Deep Learning}}
  {{Toward Data‐Driven Weather and Climate Forecasting: Approximating a
  Simple General Circulation Model With Deep Learning}}.{\BBCQ}
\newblock
\APACjournalVolNumPages{Geophysical Research Letters}{45}{22}{616--12}.
\newblock
\begin{APACrefURL}
  \url{https://onlinelibrary.wiley.com/doi/abs/10.1029/2018GL080704}
  \end{APACrefURL}
\newblock
\begin{APACrefDOI} \doi{10.1029/2018GL080704} \end{APACrefDOI}
\PrintBackRefs{\CurrentBib}

\bibitem [\protect \citeauthoryear {%
Scher%
\ \BBA {} Messori%
}{%
Scher%
\ \BBA {} Messori%
}{%
{\protect \APACyear {2019}}%
}]{%
Scher2019}
\APACinsertmetastar {%
Scher2019}%
\begin{APACrefauthors}%
Scher, S.%
\BCBT {}\ \BBA {} Messori, G.%
\end{APACrefauthors}%
\unskip\
\newblock
\APACrefYearMonthDay{2019}{6}{}.
\newblock
{\BBOQ}\APACrefatitle {{Generalization properties of neural networks trained on
  Lorenzsystems}} {{Generalization properties of neural networks trained on
  Lorenzsystems}}.{\BBCQ}
\newblock
\APACjournalVolNumPages{Nonlinear Processes in Geophysics
  Discussions}{}{}{1--19}.
\newblock
\begin{APACrefURL}
  \url{https://www.nonlin-processes-geophys-discuss.net/npg-2019-23/}
  \end{APACrefURL}
\newblock
\begin{APACrefDOI} \doi{10.5194/npg-2019-23} \end{APACrefDOI}
\PrintBackRefs{\CurrentBib}

\bibitem [\protect \citeauthoryear {%
Simonyan%
, Vedaldi%
\BCBL {}\ \BBA {} Zisserman%
}{%
Simonyan%
\ \protect \BOthers {.}}{%
{\protect \APACyear {2013}}%
}]{%
Simonyan2013}
\APACinsertmetastar {%
Simonyan2013}%
\begin{APACrefauthors}%
Simonyan, K.%
, Vedaldi, A.%
\BCBL {}\ \BBA {} Zisserman, A.%
\end{APACrefauthors}%
\unskip\
\newblock
\APACrefYearMonthDay{2013}{12}{}.
\newblock
{\BBOQ}\APACrefatitle {{Deep Inside Convolutional Networks: Visualising Image
  Classification Models and Saliency Maps}} {{Deep Inside Convolutional
  Networks: Visualising Image Classification Models and Saliency Maps}}.{\BBCQ}
\newblock
\APACjournalVolNumPages{2nd International Conference on Learning
  Representations, ICLR 2014 - Workshop Track Proceedings}{}{}{}.
\newblock
\begin{APACrefURL} \url{http://arxiv.org/abs/1312.6034} \end{APACrefURL}
\PrintBackRefs{\CurrentBib}

\bibitem [\protect \citeauthoryear {%
S{\o}nderby%
\ \protect \BOthers {.}}{%
S{\o}nderby%
\ \protect \BOthers {.}}{%
{\protect \APACyear {2020}}%
}]{%
Sonderby2020}
\APACinsertmetastar {%
Sonderby2020}%
\begin{APACrefauthors}%
S{\o}nderby, C\BPBI K.%
, Espeholt, L.%
, Heek, J.%
, Dehghani, M.%
, Oliver, A.%
, Salimans, T.%
\BDBL {}Kalchbrenner, N.%
\end{APACrefauthors}%
\unskip\
\newblock
\APACrefYearMonthDay{2020}{3}{}.
\newblock
{\BBOQ}\APACrefatitle {{MetNet: A Neural Weather Model for Precipitation
  Forecasting}} {{MetNet: A Neural Weather Model for Precipitation
  Forecasting}}.{\BBCQ}
\newblock
\begin{APACrefURL} \url{http://arxiv.org/abs/2003.12140} \end{APACrefURL}
\PrintBackRefs{\CurrentBib}

\bibitem [\protect \citeauthoryear {%
Taillardat%
, Mestre%
, Zamo%
\BCBL {}\ \BBA {} Naveau%
}{%
Taillardat%
\ \protect \BOthers {.}}{%
{\protect \APACyear {2016}}%
}]{%
Taillardat2016}
\APACinsertmetastar {%
Taillardat2016}%
\begin{APACrefauthors}%
Taillardat, M.%
, Mestre, O.%
, Zamo, M.%
\BCBL {}\ \BBA {} Naveau, P.%
\end{APACrefauthors}%
\unskip\
\newblock
\APACrefYearMonthDay{2016}{3}{}.
\newblock
{\BBOQ}\APACrefatitle {{Calibrated Ensemble Forecasts using Quantile Regression
  Forests and Ensemble Model Output Statistics.}} {{Calibrated Ensemble
  Forecasts using Quantile Regression Forests and Ensemble Model Output
  Statistics.}}{\BBCQ}
\newblock
\APACjournalVolNumPages{Monthly Weather Review}{}{}{160301131220006}.
\newblock
\begin{APACrefURL}
  \url{http://journals.ametsoc.org/doi/abs/10.1175/MWR-D-15-0260.1?af=R}
  \end{APACrefURL}
\newblock
\begin{APACrefDOI} \doi{10.1175/MWR-D-15-0260.1} \end{APACrefDOI}
\PrintBackRefs{\CurrentBib}

\bibitem [\protect \citeauthoryear {%
Weyn%
, Durran%
\BCBL {}\ \BBA {} Caruana%
}{%
Weyn%
\ \protect \BOthers {.}}{%
{\protect \APACyear {2019}}%
}]{%
Weyn2019}
\APACinsertmetastar {%
Weyn2019}%
\begin{APACrefauthors}%
Weyn, J\BPBI A.%
, Durran, D\BPBI R.%
\BCBL {}\ \BBA {} Caruana, R.%
\end{APACrefauthors}%
\unskip\
\newblock
\APACrefYearMonthDay{2019}{7}{}.
\newblock
{\BBOQ}\APACrefatitle {{Can machines learn to predict weather? Using deep
  learning to predict gridded 500‐hPa geopotential height from historical
  weather data}} {{Can machines learn to predict weather? Using deep learning
  to predict gridded 500‐hPa geopotential height from historical weather
  data}}.{\BBCQ}
\newblock
\APACjournalVolNumPages{Journal of Advances in Modeling Earth
  Systems}{}{}{2019MS001705}.
\newblock
\begin{APACrefURL}
  \url{https://onlinelibrary.wiley.com/doi/abs/10.1029/2019MS001705}
  \end{APACrefURL}
\newblock
\begin{APACrefDOI} \doi{10.1029/2019MS001705} \end{APACrefDOI}
\PrintBackRefs{\CurrentBib}

\bibitem [\protect \citeauthoryear {%
Weyn%
, Durran%
\BCBL {}\ \BBA {} Caruana%
}{%
Weyn%
\ \protect \BOthers {.}}{%
{\protect \APACyear {2020}}%
}]{%
Weyn2020}
\APACinsertmetastar {%
Weyn2020}%
\begin{APACrefauthors}%
Weyn, J\BPBI A.%
, Durran, D\BPBI R.%
\BCBL {}\ \BBA {} Caruana, R.%
\end{APACrefauthors}%
\unskip\
\newblock
\APACrefYearMonthDay{2020}{}{}.
\newblock
{\BBOQ}\APACrefatitle {{Improving data-driven global weather prediction using
  deep convolutional neural networks on a cubed sphere}} {{Improving
  data-driven global weather prediction using deep convolutional neural
  networks on a cubed sphere}}.{\BBCQ}
\newblock

\newblock
\begin{APACrefDOI} \doi{10.1002/ESSOAR.10502543.1} \end{APACrefDOI}
\PrintBackRefs{\CurrentBib}

\bibitem [\protect \citeauthoryear {%
Zhang%
, Bei%
, Rotunno%
, Snyder%
\BCBL {}\ \BBA {} Epifanio%
}{%
Zhang%
\ \protect \BOthers {.}}{%
{\protect \APACyear {2007}}%
}]{%
Zhang2007}
\APACinsertmetastar {%
Zhang2007}%
\begin{APACrefauthors}%
Zhang, F.%
, Bei, N.%
, Rotunno, R.%
, Snyder, C.%
\BCBL {}\ \BBA {} Epifanio, C\BPBI C.%
\end{APACrefauthors}%
\unskip\
\newblock
\APACrefYearMonthDay{2007}{}{}.
\newblock
{\BBOQ}\APACrefatitle {{Mesoscale predictability of moist baroclinic waves:
  Convection-permitting experiments and multistage error growth dynamics}}
  {{Mesoscale predictability of moist baroclinic waves: Convection-permitting
  experiments and multistage error growth dynamics}}.{\BBCQ}
\newblock
\APACjournalVolNumPages{Journal of the Atmospheric
  Sciences}{64}{10}{3579--3594}.
\PrintBackRefs{\CurrentBib}

\end{thebibliography}

%
%
%
%
%

\end{document}